%% file: main.tex
\documentclass[english,10pt]{DLRK_v1.1}

\usepackage{algorithm}
\usepackage{algpseudocode}
\usepackage{algcompatible}
\usepackage[capitalise]{cleveref}
\Crefname{figure}{Fig.}{Figs.}
\Crefname{table}{Tab.}{Tabs.}
\usepackage{siunitx}

\algnewcommand\algorithmicreturn{\textbf{return}}
\algnewcommand\RETURN{\State \algorithmicreturn}%

\newcommand{\PaperTitle}{\textbf{Generative Aerodynamic Design with Diffusion Probabilistic Models}}

\newcommand{\PaperAuthors}{
  \texorpdfstring
    {T. Wagenaar\ssymbol{1}, S. Mancini\ssymbol{1}, A. Mateo-Gab\'in\ssymbol{2}} 
    {T. Wagenaar, S. Mancini, A. Mateo-Gab\'in}                                  
}

\newcommand{\ContaktEmailAddress}{thomas.wagenaar@airbus.com}

\newcommand{\PaperAuthorsAffiliations}{
\ssymbol{1}\ Airbus Defence and Space GmbH, Flight Physics, Rechliner Straße, 85077 Manching, Germany\\
\ssymbol{2}\ Airbus Defence and Space SAU, Flight Physics, Paseo de John Lennon, 28906 Getafe, Spain\\
} 

\hypersetup{
    pdftitle={\PaperTitle},
    pdfauthor={\PaperAuthors},
}
\usepackage{amsmath}

\title{\PaperTitle}
\author{\PaperAuthors\vspace{0.5ex}\\\PaperAuthorsAffiliations}

\newcommand{\makeabstract}{\begin{onecolabstract}
The optimization of geometries for aerodynamic design often relies on a large number of expensive simulations to evaluate and iteratively improve the geometries. It is possible to reduce the number of simulations by providing a starting geometry that has properties close to the desired requirements, often in terms of lift and drag, aerodynamic moments and surface areas. We show that generative models have the potential to provide such starting geometries by generalizing geometries over a large dataset of simulations. In particular, we leverage diffusion probabilistic models trained on XFOIL simulations to synthesize two-dimensional airfoil geometries conditioned on given aerodynamic features and constraints. The airfoils are parameterized with Bernstein polynomials, ensuring smoothness of the generated designs. We show that the models are able to generate diverse candidate designs for identical requirements and constraints, effectively exploring the design space to provide multiple starting points to optimization procedures. However, the quality of the candidate designs depends on the distribution of the simulated designs in the dataset. Importantly, the geometries in this dataset must satisfy other requirements and constraints that are not used in conditioning of the diffusion model, to ensure that the generated geometries are physical.
\end{onecolabstract}
}

\newcommand{\paperkeywords}{Aerodynamic shape design; Generative modeling; Deep learning}

\makenomenclature

\renewcommand{\nomgroup}[1]{%
\ifthenelse{\equal{#1}{L}}{\pagebreak[2]\item[\textbf{Symbols}]}{\linebreak
  \ifthenelse{\equal{#1}{S}}{\pagebreak[2]\item[\textbf{Subscripts}]}{
    \ifthenelse{\equal{#1}{Z}}{\pagebreak[2]\item[\textbf{Abbreviations}]}{
}}}}

\begin{document}
\twocolumn[
\begin{@twocolumnfalse}
\maketitle
\vspace{0.3cm}

\makeabstract
\vspace{0.3cm}

\makepaperkeywords
\vspace{0.6cm}
\end{@twocolumnfalse}
]

\input{SymbolsAndAbbreviations} 

\printnomenclature[1.7cm] 

\section{Introduction}\label{sec:introduction}
Aerodynamic shape optimization often involves a large design space, which allows for more detailed and thus more efficient final designs. At the same time, larger design spaces are more tedious to explore because evaluations of designs rely on computationally expensive Computational Fluid Dynamics (CFD) simulations. As exploring the entire design space is practically infeasible, both gradient-based and gradient-free optimization methods rely on repetitive simulations of intermediate designs to iteratively improve aerodynamic shapes \cite{Martins2022}. The number of simulations is only increased when additional disciplines besides aerodynamics are considered in a multi-disciplinary optimization procedure, as this might lead to evolving aerodynamic requirements and constraints due to its multi-objective nature.

Fortunately, recent advances in machine learning methods promise to accelerate aerodynamic shape optimization methods \cite{Li2022}. One class of methods aims to create a reduced representation of the design space, also termed a latent space. The rationale is that most combinations of parameters lead to nonphysical designs, so that only a portion of the design space is useful. For example, it is possible to deduce a compressed design space from a set of existing feasible designs \cite{Chen2020, Wang2023}. Another class of methods rather focuses on reducing the computational cost of shape evaluations, for example by building surrogate models of high-fidelity simulations \cite{Sun2019}. Both these methods have successfully been combined, leading to rapid aerodynamic shape optimization \cite{Du2021}.

One common caveat of these methods is that they do not modify the iterative nature of the optimization procedure. This not only means that optimization procedures must be repeated whenever requirements change, but iterative optimization procedures are often sensitive to starting conditions and the definition of the objective function, and therefore convergence is not necessarily guaranteed. An approach that promises to alleviate these issues is deep generative design, which uses machine learning methods that distill information from previous designs to directly infer new designs \cite{Oh2019}. Depending on the accuracy of the design, it can be used as a final design or as the starting point for the traditional optimization procedure. 

Deep generative design has already been applied to aerodynamic shape optimization, in particular to two-dimensional airfoils. For example, Achour et al. \cite{achour_development_2020} and Tan et al. \cite{tan_airfoil_2022} use conditional Generative Adversarial Networks (GANs) to generate airfoils based on desired performance classes. However, the generated airfoils are not always smooth. Due to the sensitivity of aerodynamics to sharp features, this is problematic and Chattoraj et al. \cite{chattoraj_tailoring_2024} reduced this issue by adding an additional smoothness objective. Since only a handful of performance classes are considered to start the generation, other authors  extend the methods to work with full polars \cite{yilmaz_conditional_2020} or ranges \cite{Nobari2021}.

While using GANs for deep generative aerodynamic shape design is promising, their training procedure is unstable and therefore tedious \cite{Wiatrak2019}. In the field of image generation, from which GANs originate, they have already been superseded by Diffusion Probabilistic Models (DPMs) \cite{Dhariwal2021}. Recently, the advantages of DPMs compared to GANs have also already been established for topology optimization \cite{Maz2023}. Therefore, it seems beneficial to investigate how well DPMs perform on deep generative aerodynamic design.

While the usage of DPMs for geometry generation has already been investigated in several fields, for example to make a general framework \cite{Zhao2024}, to generate molecules \cite{Morehead2024} and to generate various objects \cite{Zeng2022}, their application to aerodynamics remains limited. To the authors' best knowledge, their only application to aerodynamic shape optimization so far still relies on an iterative optimization procedure to generate airfoils \cite{Diniz2024}. 

Therefore, the objective of this paper is to show that DPMs can be used for deep generative aerodynamic shape design. In addition, we will further formalize the deep generative aerodynamic shape design approach. Firstly, we will use a specific geometry parameterization so that the shapes are inherently smooth. Secondly, we provide a systematic way of conditioning the models to take into account both design constraints and requirements. Thirdly, we will show that diverse candidate shapes can be generated for the same operating conditions. Lastly, we will show that the generated shapes are generally different than those in the training dataset. 

\section{Methodology}
As an example, the generation of two-dimensional airfoils is considered. In particular, the task is to generate airfoil geometries that have a certain lift coefficient $C_L$, drag coefficient $C_D$ and moment coefficient $C_M$ at a specified angle of attack. While this example is rather simple, the methodology can be extended to more complex cases such as those considering the maximum lift $C_{L,\mathrm{max}}$, lift-to-drag ratio $L/D$ or even an entire lift polar similar to Nobari et al. \cite{Nobari2021}. 

\subsection{Diffusion probabilistic models}\label{sec:diffusion}
Diffusion models are originally inspired by non-equilibrium thermodynamics \cite{Dickstein2015}, but were later successfully applied to image generation \cite{Ho2020}. They operate through two processes, the first being the so called \textit{forward process}. Here, samples $\mathbf{x}$ of the dataset are taken and Gaussian noise is added in steps. If at each time step $t$ the variance of the noise added is $\beta_t$, the noised samples $\mathbf{x}_t$ can be directly determined in the closed form given in \cref{eq:forward}:

\begin{equation}
    \mathbf{x}_t = \sqrt{\bar{\alpha}_t} \mathbf{x}_0 + \sqrt{1-\bar{\alpha}_t}\bm{\epsilon}
    \label{eq:forward}
\end{equation}

Here $\bm{\epsilon} \sim \mathcal{N}(\mathbf{0}, \mathbf{I})$ is Gaussian noise, $\alpha_t=1-\beta_t$ and $\bar{\alpha}_t = \prod_{s=1}^t \alpha_s$. At the end of the forward process, at some time step $t=T$, the samples $\mathbf{x}_T$ are diffused so much that they are essentially Gaussian noise. When the original samples are parameters describing feasible airfoils, the final samples represent highly deformed airfoils that do not resemble the original airfoils anymore. 

\begin{algorithm}
\caption{Sampling procedure \cite{Ho2020}}\label{alg:reverse}
\begin{algorithmic}[1]
\STATE $\mathbf{x}_T \sim \mathcal{N}(\mathbf{0}, \mathbf{I})$;
\FOR{$t = T,...,1$}{ \State $\mathbf{z}\sim \mathcal{N}(\mathbf{0},\mathbf{I})$ if $t>1$, else $\mathbf{z}=\mathbf{0}$
\State $\mathbf{x}_{t-1} = \frac{1}{\sqrt{\alpha_t}}\left(\mathbf{x}_t - \frac{1-\alpha_t}{\sqrt{1-\bar{\alpha}_t}}\bm{\epsilon}_\mathbf{\theta}(\mathbf{x}_t,t)\right) + \sqrt{\beta}_t \mathbf{z}$ \ENDFOR}\\
\textbf{return} $\mathbf{x}_0$
\end{algorithmic}
\end{algorithm}

The fundamental aspect of DPMs is however in the \textit{reverse} or \textit{denoising process}, where a neural network is used to estimate the noise $\bm{\epsilon}$ based on a noisy sample $\mathbf{x}_t$. If this estimation is accurate, it allows to generate samples similar to those in the training dataset according to \cref{alg:reverse}. In other words, this allows to generate airfoils that are similarly distributed but not necessarily identical to the airfoils in the dataset. Since this process is stochastic, multiple generated airfoils will not necessarily be the same even if the same starting noisy sample $\mathbf{x}_T$ is used. This is an advantage compared to existing deterministic generative methods, which are only able to generate a single airfoil for any given set of features.

\begin{algorithm}
\caption{Training procedure \cite{Ho2020}}\label{alg:training}
\begin{algorithmic}[1]
\STATE \textbf{repeat}
\STATE \ \ \ \ $\mathbf{x}_0 \sim q(\mathbf{x}_0)$
\STATE \ \ \ \ $t \sim \mathrm{Uniform}({1,...,T})$
\STATE \ \ \ \ $\bm{\epsilon}\sim \mathcal{N}(\mathbf{0}, \mathbf{I})$
\STATE \ \ \ \ Take gradient descent step on \newline \phantom{tes}$\nabla_\theta ||\bm{\epsilon} - \bm{\epsilon}_\theta (\sqrt{\bar{\alpha}_t} \mathbf{x}_0 + \sqrt{1-\bar{\alpha}_t}\bm{\epsilon},t)||^2$
\STATE \textbf{until} converged
\end{algorithmic}
\end{algorithm}

To obtain an accurate noise estimation $\bm{\epsilon}_\theta$, a neural network with parameters $\theta$ is trained using the procedure outlined in \cref{alg:training}. It involves taking a sample from the dataset $q$, calculating the noisy sample at some time step $t$, and then comparing the estimated noise for this sample to the actual noise. Gradient descent is then applied to the neural network parameters based on the difference between the two, slowly improving the estimation. 

While generating new airfoils can be useful on its own, the current procedure does not yet allow for the generation of airfoils with certain features. One way to achieve this is to train multiple models, each trained on a certain subset of the dataset \cite{achour_development_2020}. However, a more general method to achieve \textit{conditional diffusion} is to add (a representation of) the features $\mathbf{f}$ of the airfoil as additional inputs to the neural network, in other words $\bm{\epsilon}_\theta=\bm{\epsilon}_\theta (\mathbf{x}_t, t, \mathbf{f})$. In the considered example, $\mathbf{f}~=~(C_L, C_D, C_M)$ so that the sampling procedure in \cref{alg:reverse} will generate airfoils with the given lift, drag and moment coefficients.

\subsection{Airfoil parameterization}\label{sec:parameterization}
\begin{figure}[H]
    \centering
    \begin{subfigure}[b]{\columnwidth}
        \centering
        \includegraphics[width=\columnwidth]{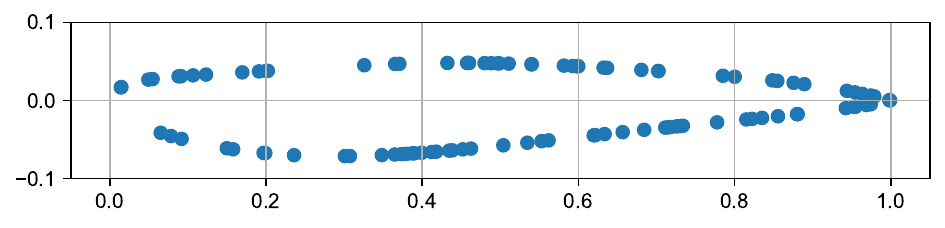}
        \caption{Point cloud}
    \end{subfigure}
    \begin{subfigure}[b]{\columnwidth}
        \centering
        \includegraphics[width=\columnwidth]{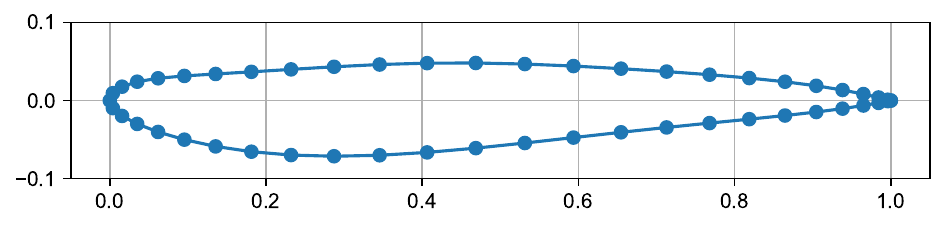}
        \caption{Mesh}
    \end{subfigure}
    \begin{subfigure}[b]{\columnwidth}
        \centering
        \includegraphics[width=\columnwidth]{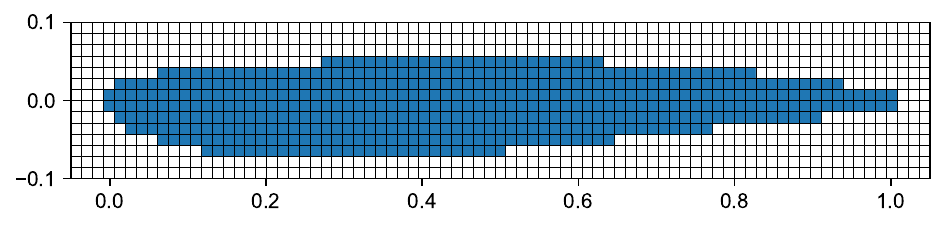}
        \caption{Voxel grid}
    \end{subfigure}
    \begin{subfigure}[b]{\columnwidth}
        \centering
        \includegraphics[width=\columnwidth]{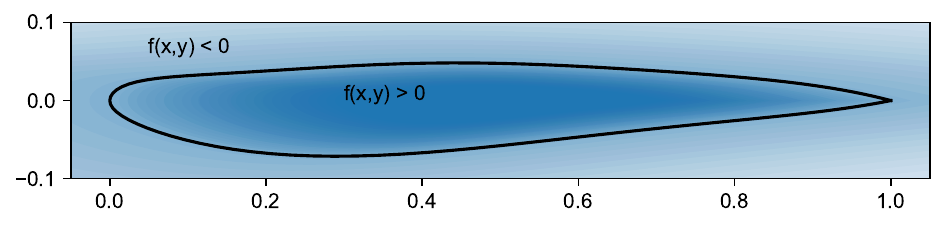}
        \caption{Signed distance function}
    \end{subfigure}
    \caption{Different geometry parameterizations applied to two-dimensional airfoils.}
    \label{fig:parameterization}
\end{figure}

An important aspect of the generation process is how the airfoils are represented by the parameters $\mathbf{x}$. Although the denoising process results in airfoils similar to those in the dataset, the parameterization can still affect how well the model can be trained and how physical the resulting airfoils are. For instance, previous works on generative models for geometry generation have used point clouds~\cite{Luo2021}, voxel grids~\cite{Smith2017}, meshes~\cite{Liu2023} and signed distance functions~\cite{Chou2023} to represent geometries. A visualization of these parameterizations applied to two-dimensional airfoils is shown in \cref{fig:parameterization}. The first three are not continuous, so that the resulting geometries only have a finite resolution. In addition, all these methods are not inherently smooth, so that small differences in the shape can lead to drastically different aerodynamic characteristics.

Therefore, an inherently smooth and continuous parameterization is used based on Bernstein polynomials \cite{kulfan2008}. Apart from these two properties, this parameterization also allows to enforce "airfoil-like" geometries that for example have rounded leading edges and sharp trailing edges. Both the upper and lower surface of the airfoil are modelled according to \cref{eq:geometry}, where $N_1$ and $N_2$ are general airfoil shape parameters, $c$ is the chord and $B_n$ is a Bernstein polynomial of order $n$. 

\begin{equation}\label{eq:geometry}
    \frac{y}{c} = \left(\frac{x}{c}\right)^{N_1} \left(1- \frac{x}{c}\right)^{N_2} \cdot B_n \left(\frac{x}{c}\right)
\end{equation}

The Bernstein polynomials are defined using equation \cref{eq:bernstein}, where $A_i$ are the polynomial coefficients and thus the geometry parameters of the airfoil. 

\begin{equation}\label{eq:bernstein}
    B_n \left(\frac{x}{c}\right) = \sum_{i=0}^n A_i \binom ni \left(\frac{x}{c}\right)^i \left(1-\frac{x}{c}\right)^{n-i}
\end{equation}

\begin{figure}[H]
    \centering
    \includegraphics[width=\columnwidth]{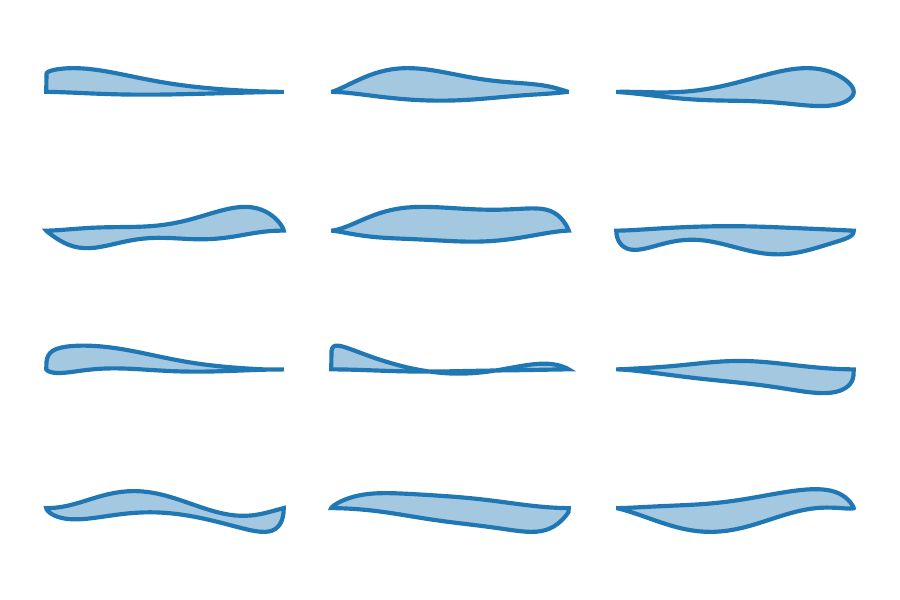}
    \caption{Randomly generated airfoils using Bernstein polynomials with $\bm{n=6}$ coefficients and random values for the shape parameters $\bm{N_1}$ and $\bm{N_2}$.}
    \label{fig:airfoils}
\end{figure}

In other words, the samples considered in \cref{sec:diffusion} take the form of \cref{eq:sample} where $A_l$ and $A_u$ represent the parameters for the lower side and upper side of the airfoil respectively. Examples of generated airfoils using this parameterization are shown in \cref{fig:airfoils}. While the geometries are certainly airfoil-like, some geometries are self-intersecting, have sharp leading edges or round trailing edges.

\begin{equation}\label{eq:sample}
    \mathbf{x} = \left\{A_{l,0}, A_{l,1}, ..., A_{u,0}, A_{u,1}, ... \right\}
\end{equation}

Nevertheless, using this representation provides smooth and continuous airfoils, with exception of the leading and trailing edges. It also significantly reduces the number of parameters that are needed to represent a geometry compared to the previously mentioned parameterizations. As mentioned in the introduction, design spaces are often unnecessarily large and can benefit from reduced representations, termed latent spaces. This is also the fundamental thought behind latent DPMs \cite{Rombach2021}, which first convert the parameter space $\mathbf{x}$ to a reduced latent space $\mathbf{\hat{x}}$. As a result, both the forward and reverse process occur in much lower-dimensional spaces which can improve the computational cost of the method.

While the Bernstein coefficients in \cref{eq:geometry} can also be considered a latent space, typically the mapping of the original space to the latent space is learned by means of an autoencoder \cite{Wang2016}. Similar techniques have been applied to generative methods for airfoil geometries~\cite{Chen2020}, but it is important to note that using a latent space does not necessarily solve the issues of continuity and smoothness. In the present work, the additional use of a latent space is not considered since the Bernstein design space is already significantly compressed. In fact, compressing the design space further might actually pose problems because DPMs rely on some concept of "noise". When the design space is too compressed, even randomly sampled sets of parameters lead to an airfoil that appears to come from the original dataset. As a result, it may become hard to distinguish the airfoils at $t=0$ and the "noisy" airfoils at $t=T$, affecting performance.

\subsection{Dataset description}\label{sec:dataset}
Before going into detail on how the dataset itself is generated, it is important to discuss the choices in filtering or conditioning features. In the considered case, conditioning only occurs for the lift, drag and moment coefficients. Any other features that are not included as conditioning variables may thus vary freely, such as the thickness or camber of the airfoil. If airfoils with unreasonable values for these features are included in the dataset, unfeasible airfoils may be generated. Therefore, it is important to filter these airfoils out of the dataset before training. Another option is to leave the dataset untouched and to simply condition the model on all relevant features, although this must be combined with optional or range-based conditioning to allow features to vary when they do not need to be constrained. However, this option is not investigated in the present work. 

\begin{figure}[H]
    \centering
    \includegraphics[width=\columnwidth]{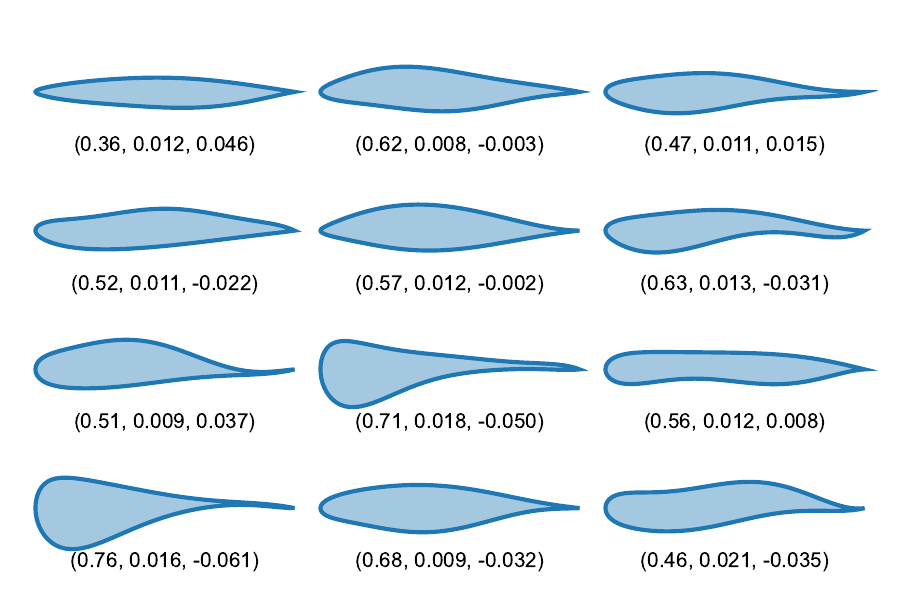}
    \caption{Subset of airfoils in the dataset along with their features ($\bm{C_L}$, $\bm{C_D}$, $\bm{C_M}$).}
    \label{fig:filtered}
\end{figure}

In general, filtering should not be done based on features that are also used to condition the model. For example, removing airfoils with negative lift coefficients from the dataset when conditioning on the lift coefficient may seem logical. However, due to the strong generalization capabilities of neural networks even "bad" geometries may provide information to the model on how to generate "good" geometries. For this same reason, the choice is made to not use the airfoils in the UIUC airfoil dataset \cite{Selig1996} as it mostly includes optimized airfoils, so that it may not include geometries matching more rarely desired features.

Instead, the dataset of airfoils is generated by sampling the parameters in \cref{eq:sample} uniformly. In particular, the airfoil shape is set to $N_1=0.5$ and $N_2=1.0$ for NACA-like airfoils with rounded leading edges and sharp trailing edges. Furthermore, Bernstein polynomials of order 6 are chosen for both the upper and lower side, resulting in a total of 11 parameters since the first lower and upper coefficient are set equal for continuous curvature at the leading edge. All parameters are sampled in the range $[-0.5, 1.5]$ while $A_{l,0}$ and $A_{u,0}$ are sampled in the range $[0.3,1.0]$ to force a minimum leading edge radius. After generation, airfoils with self-intersecting surfaces are removed. Examples of resulting airfoils are shown in \cref{fig:filtered}. 

\begin{figure}[H]
    \centering
    \includegraphics[width=\columnwidth]{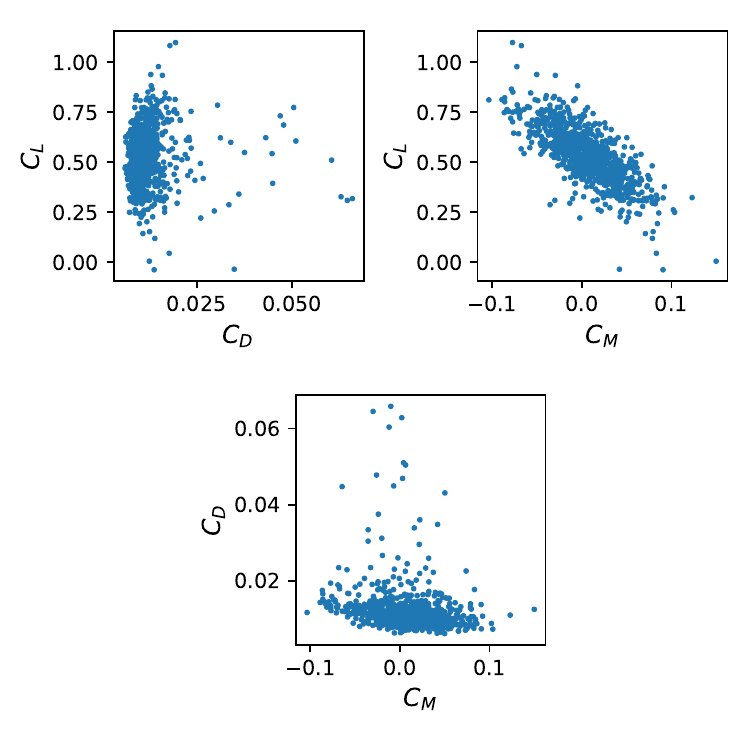}
    \caption{Pairwise distribution of the features of the airfoils in the dataset.}
    \label{fig:distribution}
\end{figure}

Afterwards, XFOIL \cite{Drela1989} is used to simulate the aerodynamics of the airfoils. Although this is a relatively low-fidelity method, the approach can be extended to higher-fidelity methods without any changes to the methodology. The airfoils are discretized to 100 cosine-spaced elements and simulated at $\mathrm{Re}=1\times 10^6$ and $\alpha=5$\degree\ using otherwise default settings. To remove outliers, the Mahalanobis distance is calculated for each point in the feature space and any points above the 99.5th distance percentile are removed. The resulting distribution of features is shown in \cref{fig:distribution}. In total, 1,000 airfoils are entered into the dataset along with their corresponding features.

\subsection{Model and training setup}
\begin{figure}[H]
    \centering
    \includegraphics[width=\columnwidth]{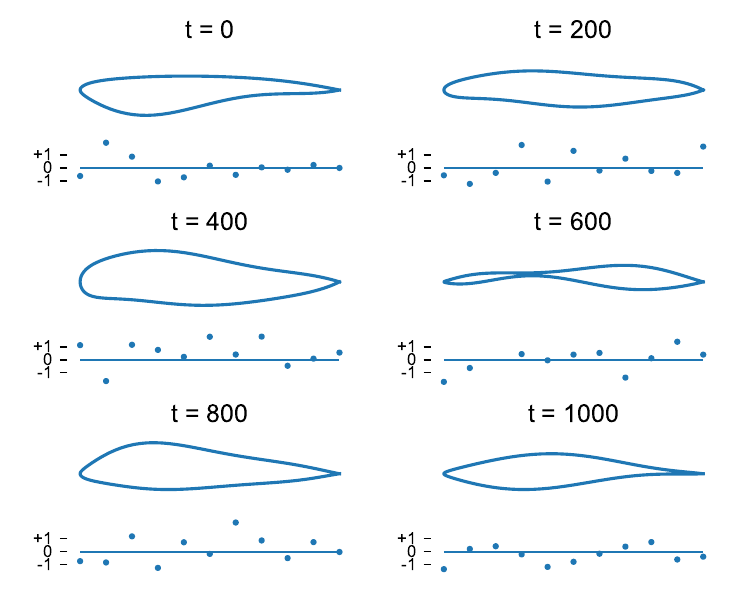}
    \caption{Forward process in the physical and latent space. The physical space is shown in terms of the geometry, while the latent space is shown with a scatter plot of the Bernstein coefficients.}
    \label{fig:forwardprocess}
\end{figure}

The dataset is split into 600 samples for training, 200 samples for validation and 200 samples for testing. Furthermore, both the samples and the features are z-score normalized based on their values in the training set. For the forward and reverse process, the variance $\beta_t$ is linearly scheduled from $1\times10^{-3}$ to $0.2$ in a total of $T=1000$ diffusion steps. A visualization of the forward process is given in \cref{fig:forwardprocess} in both the latent and physical space, confirming the assumption in \cref{sec:parameterization} that the Bernstein parameterization alone compresses the design space to an extent that the airfoil at $t=T$ is not very noisy (or out-of-distribution).

\begin{figure}[H]
    \centering
    \includegraphics[width=\columnwidth]{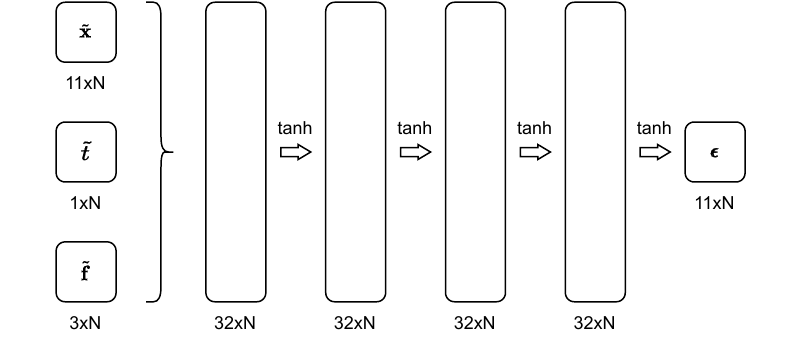}
    \caption{Architecture of the neural network used for the noise estimator $\bm{\epsilon}_{\theta}$, with $\bm{N}$ the batch size.}
    \label{fig:model}
\end{figure}

The neural network used for the noise estimator $\bm{\epsilon}_\theta~=~(\mathbf{\tilde{x}}_t, \tilde{t}, \mathbf{\tilde{f}})$ consists of 4 layers with 32 neurons each. The \texttt{tanh} activation function is used for each layer except the output layer. Here $\mathbf{\tilde{x}}$ are the normalized parameters, $\tilde{t} = t/T$ is the normalized time and $\tilde{\mathbf{f}}$ are the normalized features. Note that the noise estimator has the same output size as the number of airfoil parameters, which is 11. The neural network architecture is visualized in \cref{fig:model}.

For training, the Adam optimizer \cite{Kingma2014} is used with a learning rate of $1\times 10^{-4}$. The batch size is set to 64 and training is performed until the loss on the validation set does not improve anymore. In a nutshell, the model is trained to effectively predict the noise in the parameters given the time step and desired aerodynamic features, implicitly learning aerodynamics in the process. By iteratively subtracting this estimated noise, the corresponding geometry slowly evolves to one with the matching features.

\section{Results}
\begin{figure}[H]
    \centering
    \includegraphics[width=\columnwidth]{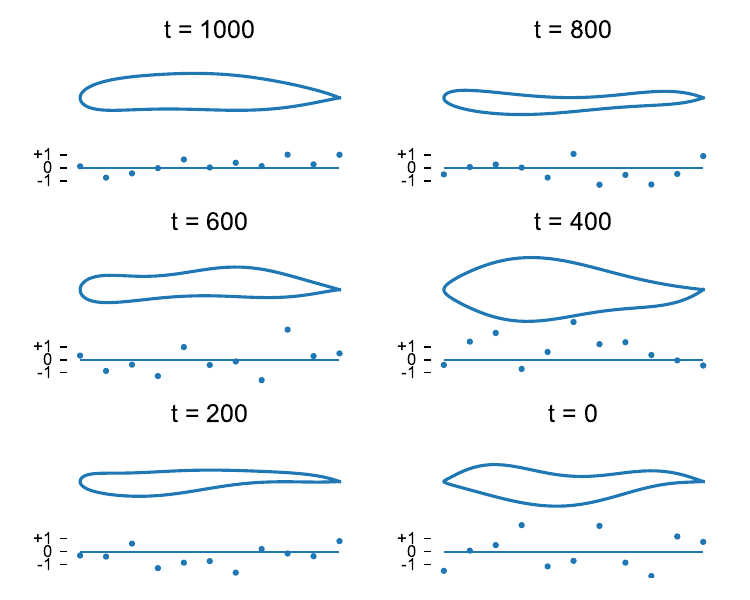}
    \caption{Reverse process in the physical and latent space.}
    \label{fig:reverse}
\end{figure}

\begin{figure}[H]
    \centering
    \includegraphics[width=\columnwidth]{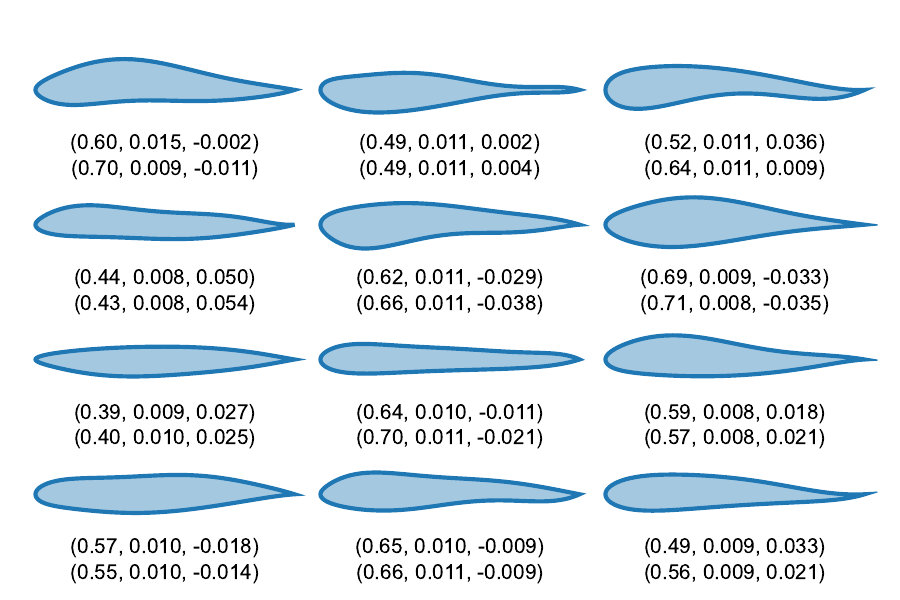}
    \caption{Examples of generated airfoils with their desired features and actual features ($\bm{C_L}$, $\bm{C_D}$, $\bm{C_M}$).}
    \label{fig:airfoilss}
\end{figure}

After training, the model can be used in the reverse process as described in \cref{alg:reverse}. An example of this process is shown in \cref{fig:reverse} and some resulting airfoils are shown in \cref{fig:airfoilss}. The generated airfoils are smooth and satisfy the requirements and constraints imposed in \cref{sec:dataset}, covering two of the four main objectives in the introduction. Nevertheless, some of the airfoils are atypical because structural constraints have not been considered. 

Two objectives still need to be verified, namely that (a) diverse candidate designs can be generated and (b) that the generated airfoils are unique and thus not identical to airfoils samples in the training dataset. But before doing so, it is important to quantify how well the aerodynamic features of the generated airfoils match the desired features.

\subsection{Feature accuracy}\label{sec:accuracy}
To quantify the accuracy of the model, the aerodynamic features of the 200 airfoils in the test set are taken and used to condition the generation of new airfoils. The resulting average Root Mean Squared Error (RMSE), Root Median Squared Error (RMdSE) and the Mean Absolute Percentage Error (MAPE) of the features are shown in \cref{tab:errors}. While the aerodynamic features of the generated airfoils are fairly accurate, there is certainly room for improvement. 

\begin{table}[H]
\caption{Errors for the different features quantified using varying metrics. MAPE for $\bm{C_M}$ is not shown because its range is zero-centered.}\label{tab:errors}
\centering
\begin{tabular}{|l|c|c|c|}
\hline
Feature    & $C_L$ & $C_D$ & $C_M$ \\ \hline
RMSE &  $8.8\times 10^{-2}$ & $8.2\times 10^{-3}$ & $1.6\times 10^{-2}$    \\ \hline
RMdSE & $3.0\times 10^{-2}$ & $9.0\times 10^{-4}$ & $7.1\times 10^{-3}$    \\ \hline
MAPE & $13.2\ \%$ & $11.1\ \%$  & - \\ \hline
\end{tabular}
\end{table}

\begin{figure}[H]
    \centering
    \begin{subfigure}{\columnwidth}
        \centering
        \includegraphics[width=\columnwidth]{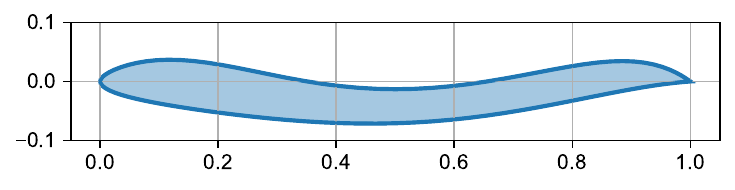}
        \caption{$\bm{C_L=0.23}$, $\bm{C_D=0.009}$ and $\bm{C_M=0.067}$}
    \end{subfigure}
    \begin{subfigure}{\columnwidth}
        \centering
        \includegraphics[width=\columnwidth]{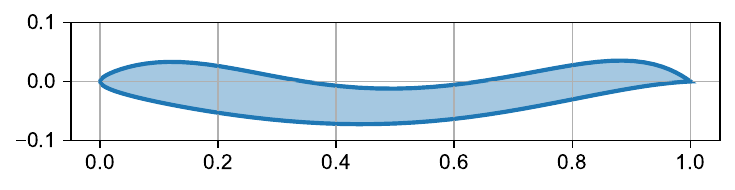}
        \caption{$\bm{C_L=0.31}$, $\bm{C_D=0.011}$ and $\bm{C_M=0.043}$}
    \end{subfigure}
    \caption{Two airfoils with similar geometries but vastly different aerodynamic features. Airfoil (a) is taken from the dataset and airfoil (b) is a slight modification of this airfoil.}
    \label{fig:twoairfoils}
\end{figure}

Note that despite the smoothness of the airfoils, the present approach still suffers from inherent aerodynamic non-linearities and modeling errors of the chosen solver. To illustrate this, consider the two airfoils in \cref{fig:twoairfoils}. While the airfoils are nearly identical, their aerodynamic features are significantly different. This may explain the large difference between the mean and median errors in \cref{tab:errors}; the model may sometimes produce a seemingly correct geometry, but due to the aerodynamic non-linearities and modeling errors, the corresponding features might deviate significantly leading to a skewed error distribution. More drastically, the errors are significantly lower when one would simply take the airfoil corresponding to the nearest neighboring point in the feature space of the training set for given desired features.

Another aspect to take into account is that feature errors may depend on the values of the features themselves, since the model is generally worse at generating airfoils for features that are less well represented in the training set. For example, the distribution in \cref{fig:distribution} shows that the density of the samples is not uniform in the feature space. In particular, the sample density for higher drag coefficients values decreases. If the errors are separated by a $C_D$ threshold as done in \cref{tab:cd}, the error in the drag coefficient increases tenfold. Therefore, it is recommended to make the dataset more uniform or to weigh the samples by their local feature density in future developments. Furthermore, note that the lift and momentum errors are less affected due to the generalization properties of neural networks.

\begin{table}[H]
\caption{Feature errors (RMSE) for different regions of drag coefficient values.}\label{tab:cd}
\centering
\begin{tabular}{|l|c|c|c|}
\hline
Feature    & $C_L$ & $C_D$ & $C_M$ \\ \hline
$C_D < 0.01$ & $6.7\times 10^{-2}$  & $9.0\times 10^{-4}$   & $1.5\times 10^{-2}$   \\ \hline
$C_D > 0.01$ & $9.4\times 10^{-2}$  & $\mathbf{9.6\times 10^{-3}}$    & $1.6\times 10^{-2}$   \\ \hline
\end{tabular}
\end{table}

Lastly, it is insightful to make a comparison to a simple method based on interpolation. Therefore, Proper Orthogonal Decomposition (POD) \cite{Berkooz1993} is applied to the training dataset to calculate its most significant modes. The resulting 11 modes are used for interpolation, which is performed by taking desired features and linearly combining these modes for each of the airfoils in the dataset.

\begin{table}[H]
\caption{Errors for the different features quantified using varying metrics for the POD approach.}\label{tab:errors3}
\centering
\begin{tabular}{|l|c|c|c|}
\hline
Feature    & $C_L$ & $C_D$ & $C_M$ \\ \hline
RMSE &  $6.3\times 10^{-2}$ & $4.1\times 10^{-3}$ & $1.1\times 10^{-2}$    \\ \hline
RMdSE & $1.8\times 10^{-2}$ & $9.0\times 10^{-4}$ & $4.2\times 10^{-3}$    \\ \hline
MAPE & $7.9\ \%$ & $12.1\ \%$  & - \\ \hline
\end{tabular}
\end{table}

\cref{tab:errors3} shows the errors on the features for the testing set. Surprisingly, the POD approach achieves lower errors in $C_L$ and $C_M$ than DPMs, while it has slightly larger errors for $C_D$. This is likely because variations in the lift and moment coefficient are largely modeled by linear models in XFOIL, while the drag coefficient depends highly on the characteristics of the boundary layer which are modeled by nonlinear, empirical rules. Therefore, DPMs likely outperform the POD approach in more non-linear regimes, for example at higher angles of attack. Nevertheless, it seems that DPMs underperform in the considered example.

\subsection{Airfoil uniqueness}
\begin{figure}[H]
    \centering
    \includegraphics[width=\columnwidth]{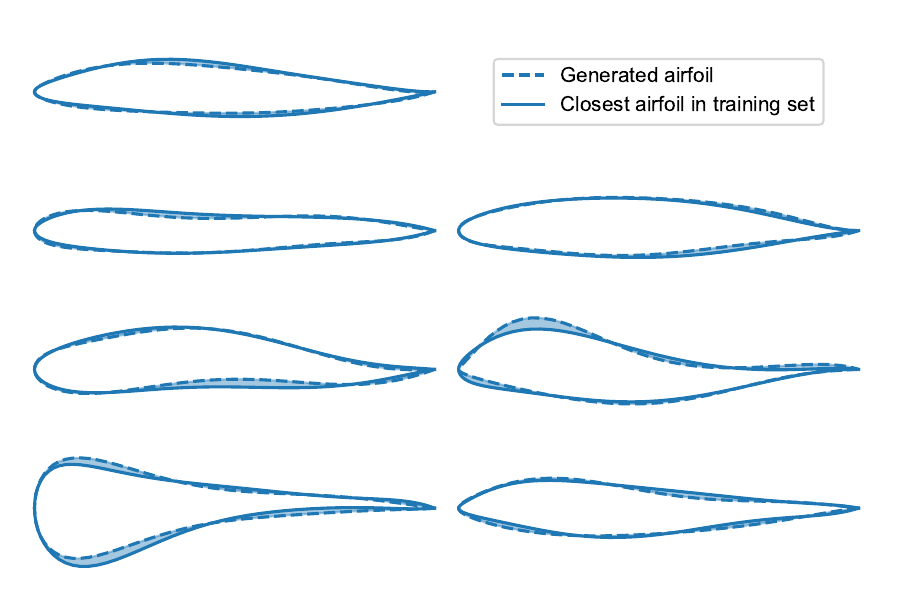}
    \caption{Examples of generated airfoils and their closest airfoils in the training set.}
    \label{fig:closest}
\end{figure}

To show that the generated airfoils from the features in the test set are unique, they are compared to the closest airfoils in the training dataset. \cref{fig:closest} shows that while the generated airfoils might resemble the types of airfoils in the training dataset, they are generally not identical. 

Note that naturally, the distance to the nearest airfoils depends on the number of airfoils and resulting feature density of the samples in the dataset. If more airfoils are included, the resemblance of generated airfoils to airfoils in the training set will increase. While this might limit the uniqueness of the generated airfoils, \cref{tab:errors2} shows that the performance of DPMs can increase beyond what POD can achieve. However, when using higher-fidelity CFD solvers to quantify the features of airfoils it is unlikely that the datasets are this large.

\begin{table}[H]
\caption{Errors for the different features when training a model with 64 neurons per layer on a training dataset consisting of 6,000 airfoils.}\label{tab:errors2}
\centering
\begin{tabular}{|l|c|c|c|}
\hline
Feature    & $C_L$ & $C_D$ & $C_M$ \\ \hline
RMSE &  $5.8\times 10^{-2}$ & $7.3\times 10^{-3}$ & $8.0\times 10^{-3}$    \\ \hline
RMdSE & $9.5\times 10^{-3}$ & $3.7\times 10^{-4}$ & $2.0\times 10^{-3}$    \\ \hline
MAPE & $6.9\ \%$ & $7.8\ \%$  & - \\ \hline
\end{tabular}
\end{table}

Regardless of the solver used, it becomes harder to achieve high feature density when the number of features increases due to the "curse of dimensionality". In the present case only three features are considered, but other features such as thickness, camber, area as well as full polars may be considered. However, neural networks in general are known to be robust to the curse of dimensionality \cite{Hu2024} and DPMs in particular are good at generalizing high-dimensional features to produce novel samples. It is therefore likely that DPMs will outperform, for instance, the POD approach when the number of features is increased even if the feature density is low. Therefore, evaluating this strategy remains as an important line of future work.

\subsection{Airfoil diversity}
An advantage of generative methods is that they can nearly instantly generate new geometries if requirements change. As mentioned in the introduction, this is especially useful in the context of multi-disciplinary optimization procedures, where requirements and constraints might evolve due to their multi-objective nature. For this purpose, it is beneficial if the model can generate diverse geometries for the same desired features because it allows to have multiple starting points for the optimization procedure. 

\begin{figure}[H]
    \centering
    \includegraphics[width=\columnwidth]{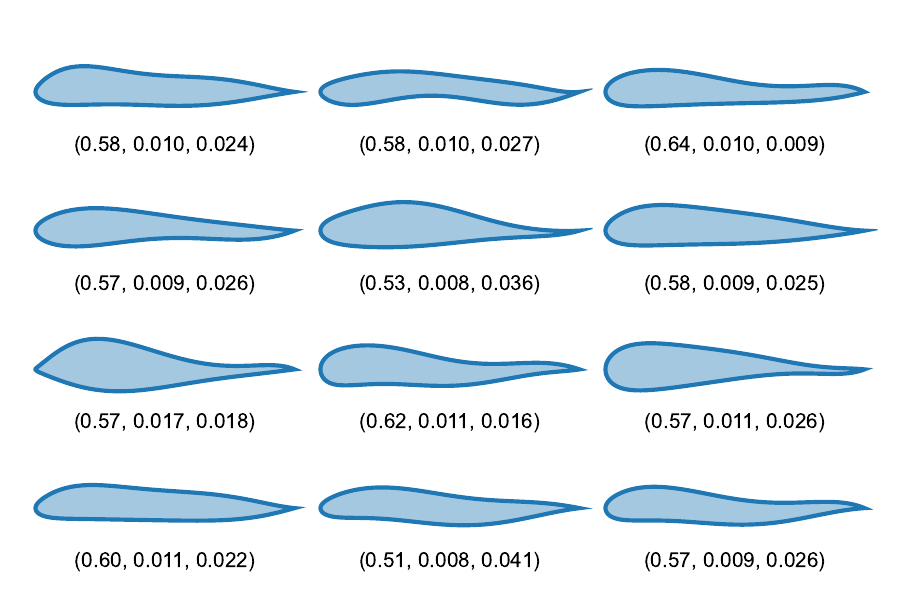}
    \caption{Generated airfoils and their features for desired features $\bm{C_L=0.6}$, $\bm{C_D=0.01}$ and $\bm{C_M=0.02}$.}
    \label{fig:diverse}
\end{figure}

\cref{fig:diverse} shows the geometries that are generated for the same set of desired features, showing a diverse set of airfoils. Note that apart from small derivations of the same airfoil, the samples also contain different "classes" of airfoils altogether. This is an advantage compared to deterministic methods, which can only output one airfoil that is often of just one class. For example, the POD approach is only able to generate one airfoil for a given set of features. While XFOIL sometimes fails to converge for both the airfoils generated by the DPM and the POD approach, the DPM can simply generate a new airfoil. For the 200 target features in the test set, the POD approach generates 15 non-converging airfoils. Nevertheless, it is important to keep in mind that the airfoils generated by the DPM do not necessarily have exactly the same features due to the inaccuracies described in \cref{sec:accuracy}.

\section{Conclusion}
The results show that diffusion probabilistic models provide an alternative method for direct aerodynamic shape design. By combining them with a field-specific latent space, they are able to generate smooth, diverse and unique airfoils. Furthermore, they allow to impose requirements and constraints easily through conditioning and filtering of the dataset. However, we acknowledge that there are certain gaps to be further explored. For example, the options for conditioning and filtering should be expanded, and in particular, future work should also consider structural requirements. In addition, the choice of a model architecture and latent space representations should be further explored. Lastly, the performance of the method should also be tested with high-fidelity solvers applied to non-linear flight conditions, as the advantages of DPMs compared to, for example, interpolation methods might become clearer. Nevertheless, the present work provides a starting point for future work on using DPMs for aerodynamic shape design.

\subsection*{\ContactSection}
\href{mailto:\ContaktEmailAddress}{\color{blue}\ContaktEmailAddress}

\subsection*{Acknowledgments}
This project has received funding from the Clean Aviation Joint Undertaking under the European Union’s Horizon Europe research and innovation program under Grant Agreement HERA (Hybrid-Electric Regional Architecture) n° 101102007. Views and opinions expressed are however those of the authors only and do not necessarily reflect those of the European Union or CAJU. Neither the European Union nor the granting authority can be held responsible.

\begin{figure}[H]
    \begin{subfigure}[b]{0.23\columnwidth}
        \includegraphics[width=\columnwidth]{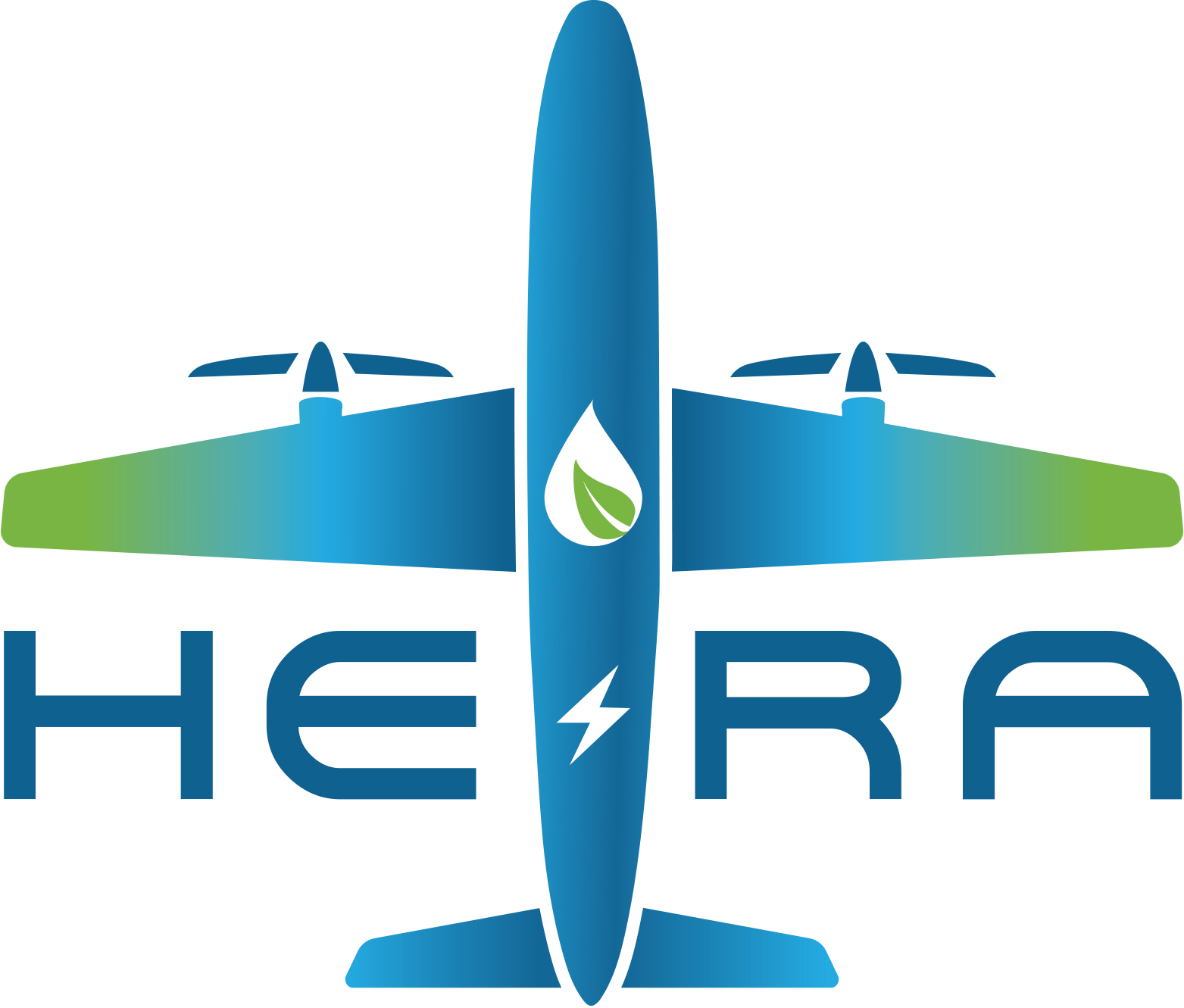}
    \end{subfigure}
    \begin{subfigure}[b]{0.76\columnwidth}
        \includegraphics[width=\columnwidth]{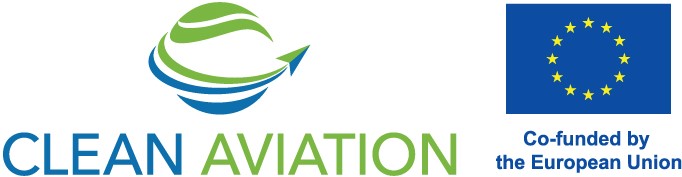}
    \end{subfigure}
\end{figure}

\bibliography{references}

\end{document}

%% file: SymbolsAndAbbreviations.tex
\nomenclature[L_alpha]{$\alpha$}{Angle of attack}
\nomenclature[L_sample]{$\mathbf{x}$}{Sample}
\nomenclature[L_time]{$t$}{Time step}
\nomenclature[L_maxtimestep]{$T$}{Maximum time step}
\nomenclature[L_noise]{$\bm{\epsilon}$}{Noise}
\nomenclature[L_variance]{$\beta$}{Variance}
\nomenclature[L_normal]{$\mathcal{N}$}{Normal distribution}
\nomenclature[L_normalsample]{$\mathbf{z}$}{Normal sample}
\nomenclature[L_identity]{$\mathbf{I}$}{Identity matrix}
\nomenclature[L_parameter]{$\theta$}{Neural network parameters}
\nomenclature[L_q]{$q$}{Dataset sampling function}
\nomenclature[L_c]{$c$}{Chord}
\nomenclature[L_N1]{$N_1$}{First shape coefficient}
\nomenclature[L_N2]{$N_2$}{Second shape coefficient}
\nomenclature[L_B]{$B$}{Bernstein polynomial}
\nomenclature[L_b]{$b$}{Bernstein basis polynomial}
\nomenclature[L_z]{$y$}{Height}
\nomenclature[L_z]{$A$}{Bernstein polynomial coefficient}
\nomenclature[L_f]{$\mathbf{f}$}{Features}
\nomenclature[L_C]{$C$}{Coefficient}
\nomenclature[L_Re]{$\mathrm{Re}$}{Reynolds number}

\nomenclature[S_l]{$L$}{lift\nomunit{}}
\nomenclature[S_d]{$D$}{drag\nomunit{}}
\nomenclature[S_g]{$M$}{moment\nomunit{}}
\nomenclature[S_d]{$l$}{lower\nomunit{}}
\nomenclature[S_g]{$u$}{upper\nomunit{}}
\nomenclature[S_g]{$\mathrm{max}$}{Maximum\nomunit{}}

\nomenclature[Z_DPM]{DPM}{Diffusion Probabilistic Model}
\nomenclature[Z_GAN]{GAN}{Generative Adversarial Network}
\nomenclature[Z_CFD]{CFD}{Computational Fluid Dynamics}
\nomenclature[Z_RMSE]{RMSE}{Root Mean Square Error}
\nomenclature[Z_RMdSE]{RMdSE}{Root Median Square Error}
\nomenclature[Z_MAPE]{MAPE}{Mean Absolute Percentage Error}
\nomenclature[Z_POD]{POD}{Proper Orthogonal Decomposition}